\documentclass[aps,pra,amssymb,amsmath,superscriptaddress,onecolumn,10pt]{revtex4-2}
\usepackage{xcolor}
\usepackage{graphicx}
\usepackage{bm}
\usepackage{times,txfonts}
\usepackage[english]{babel}
\usepackage{hyperref}
\usepackage{booktabs}
\usepackage[normalem]{ulem}
\usepackage{soul}

\def\br{\bm{\rho}}

\begin{document}

\title{Correlation Hyperspectral Imaging}

\author{Gianlorenzo Massaro}
\affiliation{Dipartimento Interateneo di Fisica, Universit\`{a} degli Studi di Bari, I-70126 Bari, Italy}
\affiliation{Istituto Nazionale di Fisica Nucleare, Sezione di Bari, I-70125 Bari, Italy}

\author{Francesco V. Pepe}\email{francesco.pepe@ba.infn.it}
\affiliation{Dipartimento Interateneo di Fisica, Universit\`{a} degli Studi di Bari, I-70126 Bari, Italy}
\affiliation{Istituto Nazionale di Fisica Nucleare, Sezione di Bari, I-70125 Bari, Italy}

\author{Milena D'Angelo}
\affiliation{Dipartimento Interateneo di Fisica, Universit\`{a} degli Studi di Bari, I-70126 Bari, Italy}
\affiliation{Istituto Nazionale di Fisica Nucleare, Sezione di Bari, I-70125 Bari, Italy}

\begin{abstract}
Hyperspectral imaging aims at providing information on both the spatial and the spectral distribution of light, with high resolution. However, state-of-the-art protocols are characterized by an intrinsic trade-off imposing to sacrifice either resolution or image acquisition speed. We address this limitation by exploiting light intensity correlations, which are shown to enable overcoming the typical downsides of traditional hyperspectral imaging techniques, both scanning and snapshot. The proposed approach also opens possibilities that are not otherwise achievable, such as sharper imaging and natural filtering of broadband spectral components that would otherwise hide the spectrum of interest. The enabled combination of high spatial and spectral resolution, high speed, and insensitivity to undesired spectral features shall lead to a paradigm change in hyperspectral imaging devices and open-up new application scenarios.
\end{abstract}

\maketitle
Spectral imaging has significantly contributed to fields such as material science, biology, and astronomy by providing detailed spatial and spectral information about the objects under study \cite{grahn2017techniques,landgrebe1999information,vane1993imaging,lu2014medical}. Spectral imaging techniques can be divided into two main categories: snapshot techniques, which are the fastest but suffer from a direct conflict between spatial and spectral resolution \cite{hagen2012snapshot,hagen2013review}, and scanning techniques, which are slower due to a trade-off between resolution (either spatial or spectral) and acquisition time \cite{grahn2017techniques,landgrebe1999information,gat2000imaging}. All traditional methods thus face limitations in resolution, sensitivity, and speed, strongly required to capture dynamic processes. A powerful alternative to conventional imaging techniques has emerged in the field of quantum imaging, where the statistical properties of light are exploited to offer new and unique advantages, such as overcoming the typical limitations of conventional devices  \cite{pittman1995optical,genovese2016real,moreau2019imaging,gilabertebasset2019perspectives,valencia2005two,gatti2004ghost,ferri2005high,scarcelli2006can,chiuri2023}. Such correlation-based imaging approaches leverage the quantum and classical correlations of photons to extract information about the object. By analyzing the statistical relationships between light beams, these methods can enhance image resolution, improve signal-to-noise ratios, and enable imaging under conditions where conventional methods fail \cite{meyers2008ghost,hardy2010ghost,chan2009twocolor,duan2019nondegenerate,karmakar2010twocolor,kalashnikov2016infrared,meyers2011turbulence,shi2012adaptive,bina2013backscattering,sun20233dcomputational,dangelo2016correlation,pepe2017diffraction,massaro2023correlated}. The underlying principle is that photons carry correlated information about the sample that can be decoded to reconstruct an image with enhanced features.

In this article, we propose a novel approach to spectral imaging, based on light intensity correlation. This technique, named correlation hyperspectral imaging (CHI), can provide detailed spectral and spatial information about the sample, overcoming some of the downsides of both snapshot and scanning techniques. Furthermore, the use of correlations has several advantages not achievable with any other technique, such as the possibility of becoming insensitive to unwanted spectral features which overlap with the spectrum of interest, and improved optical performance enabled by light spatial coherence.
These benefits make the proposed approach a promising tool for applications in biomedical imaging, where high resolution and efficiency are paramount, as well as environmental monitoring and industrial inspection, which would benefit from the loss of sensitivity to undesired spectral features, as well as by the combination of high resolution and fast acquisition.

In CHI, the combined spatial and spectral information required for spectral imaging is obtained by measuring the intensity on two separate sensors: light emitted by the sample is split into two optical arms, so that spatial information (\textit{i.e.} the panchromatic intensity distribution) is retrieved by measuring light intensity on a monochrome imaging system, while the spectrum is measured through a dedicated spectrometer, as schematically reported in Fig.~\ref{fig:scheme}.
\begin{figure}
    \centering
    \includegraphics[width=0.7\textwidth]{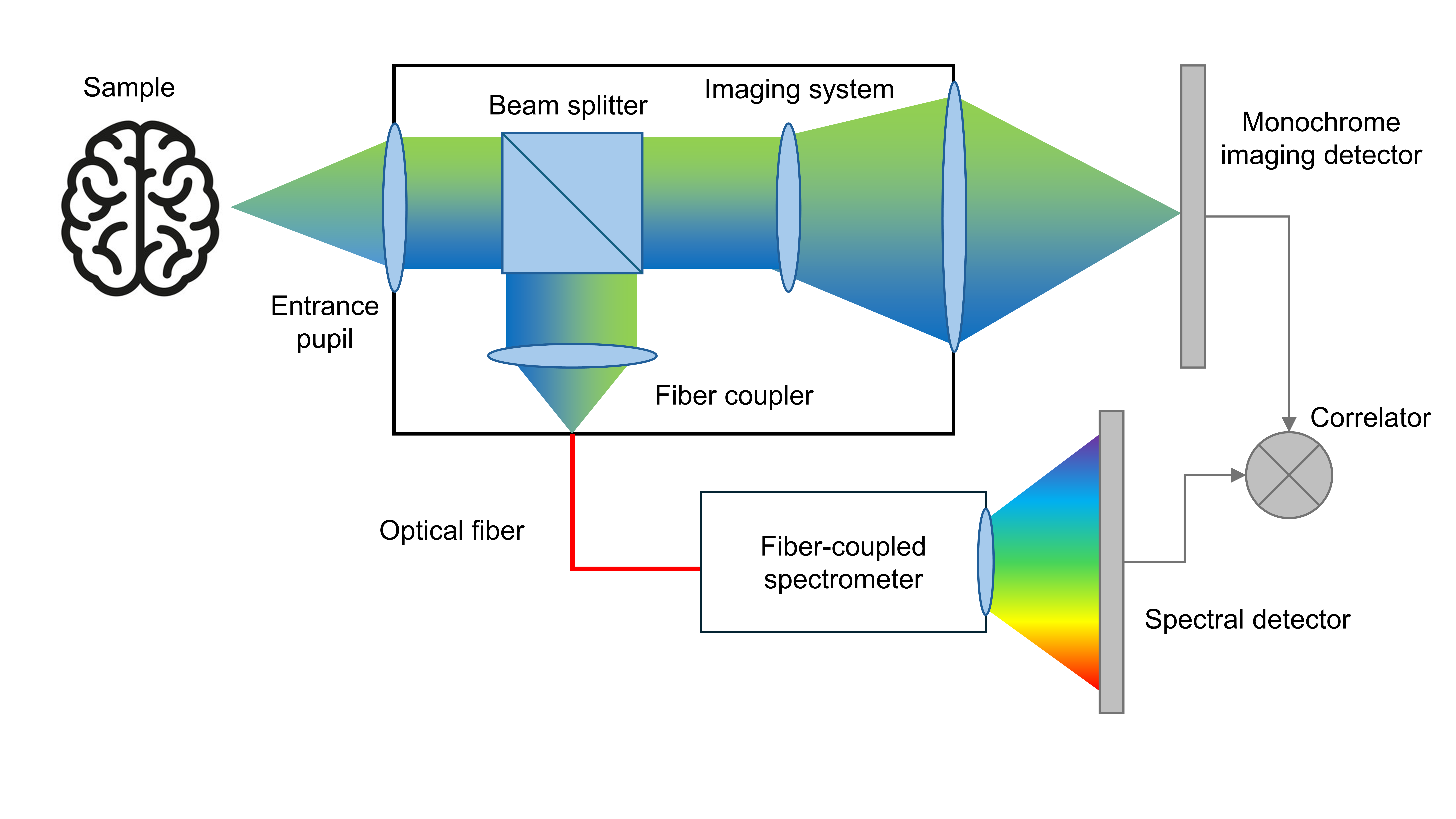}
    \caption{Schematics of the experimental setup. The intensity fluctuations detected in time by the monochrome imaging sensor and the spectrometer are correlated to obtain a hyperspectral image of the sample. }
    \label{fig:scheme}
\end{figure}
The spectral image, which combines spectrum and spatial intensity distribution, is obtained by measuring the cross-correlation function between light intensity \textit{fluctuations} at the two sensors
\begin{equation}
    \Gamma(\br_a,\br_b, \Delta t)=\left\langle
        \Delta\bar{I}_a(\br_a, t)\,\Delta\bar{I}_b(\br_b, t+\Delta t)
    \right\rangle
    \label{eq:corrFun_definition}
\end{equation}
where $\bar{I}_a$ and $\bar{I}_b$ are the 2D intensity distributions collected at the spatial ($a$) and spectral ($b$) sensors, $\Delta\bar{I}_i=\bar{I}_i-\left\langle\bar{I}_i\right\rangle$, with $i=a,b$ denotes the intensity fluctuations around its average value, $\br_a$ and $\br_b$ indicate the coordinates on the sensors surface, $t$ represents the temporal coordinate and $\Delta t$ the correlation time delay. $\left\langle X \right\rangle$ denotes the statistical average of the stochastic quantity $X$.
We shall assume the intensity form the sample to be stationary, so that the correlation function is independent of the time $t$ and only depends on the time delay $\Delta t$.

One of the key aspects of our technique is the ability to tune its spectral sensitivity to the statistical phenomena of interest. This can be seen by taking into account the finite bandwidth of the sensors, so that the quantity $\bar I_i$ appearing in \eqref{eq:corrFun_definition} is not the \textit{instantaneous} light intensity distribution, but the \textit{time-averaged} intensity measured, at time $t$, by a detector with response time $\tau_{\text{exp}}$, namely,
\begin{equation}
    \bar I_i(\br_i,t)=\frac 1{\tau_\text{exp}}\int_{t}^{t+\tau_\text{exp}}
    \left\lvert
        E_i(\br_i,z_i,t^\prime)
    \right\rvert^2 dt^\prime,
    \label{eq:frame_intensity}
\end{equation}
where $z_i$ is the axial coordinate of detector $i={a,b}$, and $E_i$ is the electric field on its surface.
The finite bandwidth of the detector is well known, in the context of correlation-based methods based on thermal light, to have detrimental effects on the measured correlation function. In fact,
This is due to the fact that, as the detector response time $\tau_\text{exp}$ becomes larger than the light coherence time, intensity fluctuations are integrated and their relative amplitude tends to vanish ($\Delta \bar I/\langle \bar{I}\rangle \to 0$), making the evaluation of correlations more and more sensitive to noise.
In the context of spectroscopy, however, one must consider that the overall spectrum is typically contributed by a plethora of different phenomena, each with its own features and bandwidth. These phenomena can mostly be considered to be \textit{statistically independent} from one another, and, as such, each one characterized by its own coherence time.
It is thus rather intuitive that, by appropriately setting the detector response time, the multi-spectral correlation function can be adjusted to loose sensitivity to broadband spectral features, having a small coherence time, and isolate fluctuations due to the narrow-band emission of the phenomena of interest.
This feature is very useful in many cases of interest for spectroscopy, where one is not interested in the known shape of broad-band contributions (\textit{e.g.} black-body spectrum), but only to the line-shaped footprint of the material of interest.

\begin{figure}[ht!]
    \centering\includegraphics[width=.6\textwidth]{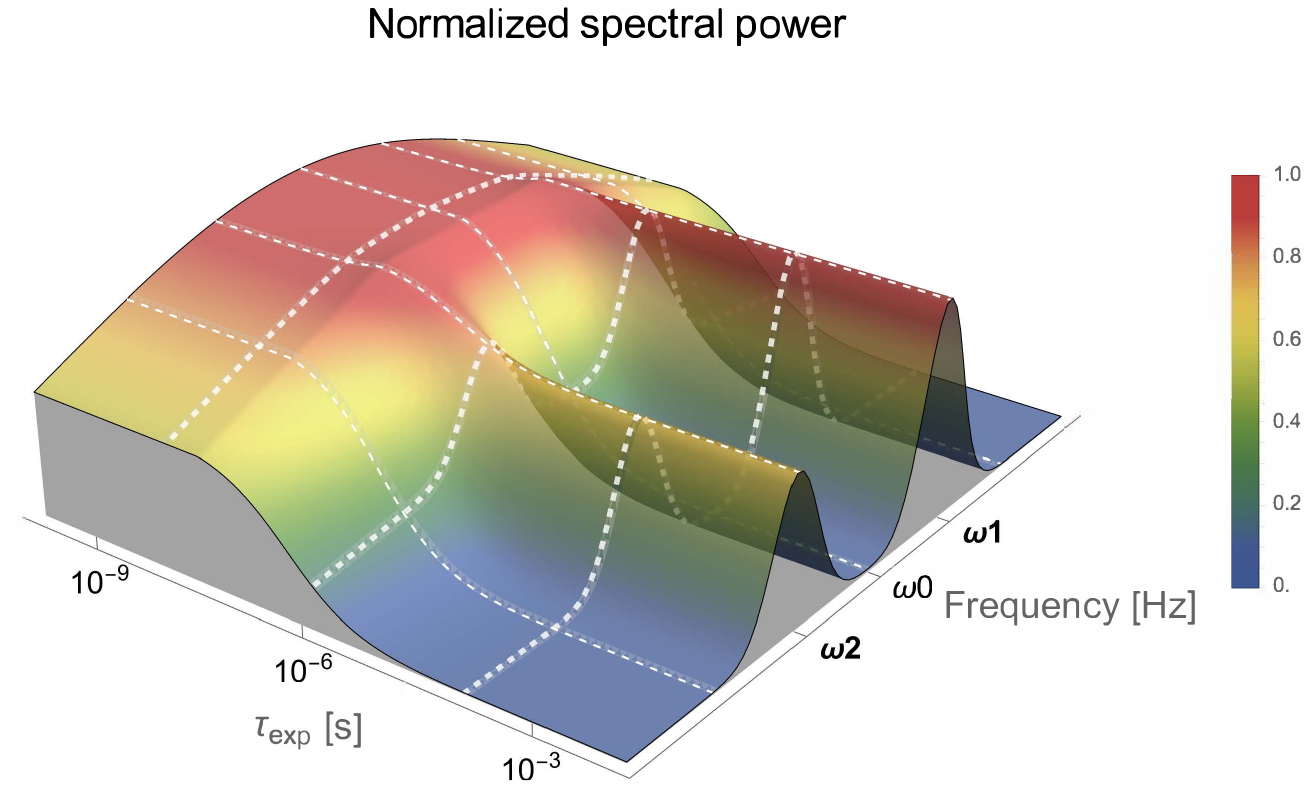}
    \caption{CHI correlation function: Improvement of the spectral contrast for varying response time $\tau_\text{exp}$, in the case of two narrow-band signals superposed to a broadband background.}
    \label{fig:spec_vs_exp}
\end{figure}

To focus on this aspect, we shall now disregard the spatial features of the sample and only consider the spectrum originating from a single spatial coordinate on the sample. In Fig.~\ref{fig:spec_vs_exp}, we report the simulated spectral part of the correlation function obtained by spanning a large range of detector response time; in the considered situation, three statistical phenomena co-exist in a small portion of the object: a broadband (low-coherence) emission, centered at a frequency $\omega_0 = 100\,$THz, characterized by a Gaussian spectrum having width $\Delta\nu_0=100\,$GHz, and two narrow-band lines centered at $\omega_{1,2} = \omega_0\pm25\,$GHz, both having bandwidth $\Delta\nu = 10\,$kHz.
The spectrum is measured through a spectrometer with resolution $\Delta\nu_\text{spec}=10\,$GHz, so that the perceived width of the narrow-line emission is defined by $\Delta\nu_\text{spec}$, although two narrow emissions can still be clearly resolved.
To better highlight the usefulness of the \textit{statistical filtering} operated by CHI through tuning of $\tau_\text{exp}$, we choose to simulate a case where the emission of photons from the broadband emission is much more likely than the emission that contributes to the two lines. We thus fixed the photon flux to be $1000$ times larger that the photon flux of the emission centered at $\omega_1$, and $1400$ times larger than the emission centered at $\omega_2$.
In these conditions, when the response time is small enough to be sensitive to the whole emission spectrum, the signal from the two lines is drowned in the uninteresting broadband background. Actually, since in the considered case $\Delta\nu\ll\Delta\nu_{\text{spec}}\ll\Delta\nu_0$, when $\tau_{\text{exp}} \ll (\Delta\nu_0)^{-1}$, the contrast between the most intense line and the broadband maximum is
\begin{equation}
    C_0\simeq
    \left(\frac{n}{n_0}\right)^2
	\left(\frac{\Delta\nu_0}{\Delta\nu_\text{spec}}\right)^2
	=10^{-4},
\end{equation}
with $n/n_0=10^{-3}$ the ratio between the photon fluxes. As clearly demonstrated in Fig.~\ref{fig:spec_vs_exp}, the increase in response time suppresses the broadband background much more than it does on the line-width emission; the highest possible contrast is obtained for $\tau_\text{exp} \gg (\Delta\nu)^{-1}$, and reads
\begin{equation}
    C_\infty\simeq
    \left(\frac{n}{n_0}\right)^2
	\frac{\Delta\nu_0}{\Delta\nu_\text{spec}}
	\frac{\Delta\nu_0}{\Delta\nu}=10^{2}.
\end{equation}
Large response times, which are typically detrimental in correlation measurements, result, in CHI, in a six-order of magnitude increase of the contrast of narrow lines over the useless broadband emission, as determined by the ratio $C_\infty/C_0=\Delta\nu_\text{spec}/\Delta\nu$.

After focusing the discussion on the characteristics of the pure spectral measurement, let us now focus on the features of the correlation function that determine the quality of the spatial image. For the sake of a lighter formalism, we shall now assume that the detectors are ideal ($\tau_\text{exp}=0$), and only consider the $x$-component of the 2D coordinates at the detector $\br_{a,b}=(x_{a,b},y_{a,b})$. The last assumption does not affect the spectral part of the measurement, since chromatic dispersion in spectrometers always occurs in one dimension (here chosen as $x_b$); it simply reduces the spatial detector to a pixel strip, limiting the analysis to \textit{linear} images. 
For the sake of simplicity, we shall further assume that the correlation time delay is fixed at $\Delta t = (z_a-z_b)/c$, which compensates for the phase shift arising from the different optical paths leading to the two sensors and maximizes the visibility of the correlation function.
In this simplified context, as shown in the Appendix, \eqref{eq:corrFun_definition} can be written as:
\begin{equation}
    \Gamma(x_a,\tilde\omega(x_b))=
    \left\lvert
        \int\int\tilde{I}(x_0,\omega)
        \,\tilde g^*_{a}(x_0,x_a)
    \tilde g_b(x_b,\omega)\,dx_0\,d\omega
    \right\rvert^2,
    \label{eq:corr_coherent}
\end{equation}
where $\tilde I(x_0,\omega)$ is the emission spectrum of the sample, as a function of the emission coordinate $x_0$ on the sample plane; the two functions $g_{a,b}$ represent the frequency-dependent point-spread function (PSF) propagating the electric field from the sample plane to the spatial and spectral detector, respectively.
Since the optical layout of the spectrometer arm is such to establish a correspondence between frequencies and detector coordinates, $x_b\to\bar{\omega}(x_b)$, we chose to make this dependence explicit in the second argument of the correlation function.
As per ~\eqref{eq:corr_coherent}, the correlation function encoding the spectral image is a continuous superposition of\textit{coherent-like} images: even though the imaging function is not sensitive to the phase of light on the sample, it still depends on the phase of the transfer functions; as recently demonstrated, such a property determines very distinctive features compared to standard \textit{incoherent} images, such as very slow loss of focus and better resolution out of focus, both in intensity-based \cite{massaro2024direct} and in correlation-based imaging \cite{pepe2017diffraction}.

To show the convenience of the coherent-like imaging capability of CHI, let us compare it with a typical frequency-sweeping technique. To this end, we shall consider the image obtained through the same imaging system of arm $a$, upon isolating a single frequency by means of a filter centered around $\omega^\prime$ and bandwidth $\delta\nu_\text{f}$; of course, the acquisition of the complete spectral image requires that the center frequency of the filter sweeps the whole spectral range of interest.
By modeling the filter frequency response through a positive function $\tilde f(\omega^\prime-\omega)$, peaked around $\omega^\prime$ and vanishing outside of the bandwidth, the obtained spectral image is:
\begin{equation}
    I_\text{sw}(x_a,\omega^\prime)=
    \int\tilde I(x_0,\omega)\,
    \left\lvert
        \tilde g_a(x_0,x_a)
    \right\rvert^2\, \tilde f(\omega^\prime-\omega)\,dx_0\,d\omega.
    \label{eq:sweep}
\end{equation}
\eqref{eq:corr_coherent} and \eqref{eq:sweep} are very similar: The imaging device described by $g_a$ and the spectral device (described by either $g_b$ or $f$) play the same role, giving rise to separate convolutions with the spatial and frequency dependence of $\tilde I$.
The similarities are even more marked by selecting a filter bandwidth that matches the bandwidth of the spectrometer, namely $f(\omega^\prime-\omega)=|\tilde g_b(x_b(\omega^\prime),\omega)|^2$.
In fact, even if the optical and spectral performance are evenly matched in the two cases under inspection, so that the same exact functions are involved in the two equations, the key difference in the image formation process remains. This is due to the fact that standard techniques rely on an \textit{incoherent} formation process, as determined by all positive-valued PSF appearing in \eqref{eq:sweep}, whereas the measurement of the correlation function results in a coherent-like superposition, as determined by the complex functions appearing in \eqref{eq:corr_coherent}.

\begin{figure}[ht!]
    \centering\includegraphics[width=.5\textwidth]{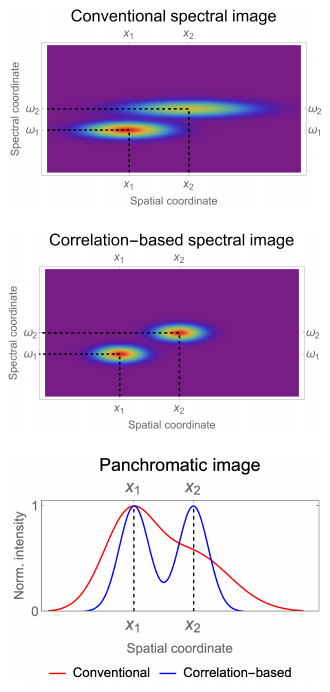}
    \caption{Comparison between the optical performance of conventional sweeping hyperspectral imaging (top) and CHI (middle). Bottom: panchromatic image obtained through integration in frequency.}
    \label{fig:comparison}
\end{figure}

This can clearly be seen in Fig.~\ref{fig:comparison}, showing the optical performance advantage of the coherent-like spectral image encoded in the correlation function of CHI, over the (incoherent) image obtained through frequency sweeping.
The simulation reports the spectral images of a sample consisting of two statistically independent emitters; both sources emit a Gaussian spectrum, centered at frequencies $\omega_1$ and $\omega_2$, respectively.
In order to compare the techniques as fairly as possible, the frequency filter and the spectrometer response have been set to be equal, Gaussian-shaped, with bandwidth $\Delta\nu_\text{spec}=\Delta\nu_f=|\omega_2-\omega_1|/\pi$
matched to $(\omega_2-\omega_1)/\pi$, where $\omega_{1,2}$.
We have chosen $\omega_1=100\,$THz and $\omega_2 = \omega_1+50\,$GHz.
Each of the two statistical emissions from the sample is characterized by a very small bandwidth, so that their spectral image is defined by the filter/spectrometer resolution.
Spatial information for the two cases is measured through the same imaging device: a 4-f system made of two identical lenses, placed at twice their focal length from each other, composed of two $50$-mm lenses; the entrance pupil has a $0.2$ numerical aperture, with Gaussian apodization for ease of calculation.
In 4-f system, a perfectly focused image is retrieved by the image sensor when the sample is in the first focal plane of the first lens. It is well known that a large-aperture (incoherent) imaging system can only retrieve a sharp image in close proximity of the focus; the image features undergo severe blurring upon moving away from the focus. On the contrary, coherent-like imaging gives rise to a much longer depth of field, regardless of the numerical aperture \cite{massaro2024direct}.
As demonstrated in Fig.~\ref{fig:comparison}, the coherent-like nature of CHI thus implies a significant advantage when the two sources emitting at $\omega_1$ and $\omega_2$ are not put in perfect focus. In the presented simulation, the two sources are identical: Gaussian-shaped, with a width of $1\,$mm, centered at $x_{1,2}=\pm 2\,$ mm, respectively, but are defocused by $+1\,$cm and $-1\,$cm, respectively (negative defocusing indicating an axial displacement towards the entrance pupil, and vice versa).
Comparison between the upper and middle panels of Fig.~\ref{fig:comparison} shows that the spectral properties of the two techniques (vertical axes) are identical, in accordance to our choice of matching the filter and spectrometer performance. However, the difference in imaging performance between the two hyperspectral imaging modalities is immediately evident: images from conventional hyperspectral imaging are much more blurred, hence less localized, than their counterparts obtained through intensity correlations in CHI.
This is particularly evident by considering the panchromatic images obtained by integrating over the whole spectrum, as reported in the bottom plot of Fig.~\ref{fig:comparison}.
We should also remark that, 
the conventional (\textit{incoherent}) information is always available in CHI: the imaging arm, alone, gives the monochromatic (\textit{incoherent}) image, and the spectral arm the emission spectrum.
Most interestingly, CHI gives a panchromatic image by integrating the correlation function over frequency; such a correlation-based image has a much larger depth of field than the correlation-free (i.e., intensity-based) image: the very different nature of the two imaging modalities keeps showing up even when spectral features are neglected.

In the example of Fig.~\ref{fig:spec_vs_exp}, we have considered a physical regime in which the condition of isochronous detection ($\Delta t = (z_a-z_b)/c$) is approximately satisfied for the whole sample, regardless of the position of details on the optical axis. 
This situation is perfectly acceptable in CHI, thanks to the phase sensitivity of the correlation function and its connected coherent-like imaging behaviour, which make CHI capable of collecting a focused image of the whole sample, even with a modest numerical aperture. On the contrary, conventional techniques based on \textit{incoherent} imaging yield a blurred image. This advantage of CHI is envisioned to be particularly useful in scenarios such as industrial inspection, when one wants to obtain a high resolution spectral image with as much depth of field as possible, disregarding the details about the precise axial placement of the sample features.

Apart from the peculiar features of CHI so far discusses, the proposed approach has the further advantage of overcoming the intrinsic limitations of both scanning and snapshot hyperspectral imaging techniques.
Snapshot techniques are based on the simultaneous measurement, on a single sensor, of both the spectrum and the image of the sample, and are typically achieved by mounting spectral filters on a conventional camera; each element of the image is thus observed at different frequencies. This approach entails a spatial resolution loss that is \textit{at least} equal to the number of measured chromatic components. For instance, a Bayer matrix typically used in RGB imaging entails a loss of resolution by a factor four.
Non-snapshot, a.k.a. scanning, techniques retrieve the spectral cube by acquiring one of its three dimensions over time: 
A single snapshot can be either a 2D image at a single frequency, so that the chromatic component is obtained by frequency-scanning (\textit{e.g.}, with a series of filters), or a linear image along with its spectrum, in which case the other axis is acquired by spatially scanning the sample along the remaining dimension.
Thanks to the use of two disjoint sensors simultaneously collecting spatial and spectral information, CHI can acquire the spectral cube without being intrinsically subject to any such trade-off, thus enabling combining high spatial and spectral resolution, with high acquisition speed.

Another disadvantage of conventional hyperspectral imaging is its intrinsically wasteful nature of photons emitted by the sample. This is true for both snapshot techniques, where most of the optical intensity is lost because of frequency filtering, and scanning techniques, where one is only sensitive to a small portion of the object or its spectrum, at any given time.
Conversely, the peculiar CHI decoupling of spatial and spectral measurement makes use of the entirety of the photons emitted by the object, possibly limited only by a finite angular acceptance of the device and optical losses.

In summary, correlation-based hyperspectral imaging represents a significant advancement in the field of spectral imaging. By utilizing the inherent statistical properties of light, CHI offers new possibilities for high-resolution, sensitive, fast and robust imaging across a wide range of applications. As research and technology in this area continue to advance, correlation-based imaging is expected to play a crucial role in addressing some of the most challenging problems of both shapshot and scanning conventional methods.

\begin{acknowledgments}
\textit{Acknowledgments.---} M.D. and F.V.P. acknowledge support by PNRR MUR project PE0000023-NQSTI (National Quantum Science and Technology Institute). The activity is synergic to projects QuaASiModo, financed the MUR Dipartimenti di Eccellenza '23-'27, and QUISS, financed by INFN. 
\end{acknowledgments}

\appendix

\section{Theoretical model}

Eq.~(1) and~Eq.~(2) of the main text report, respectively, the definition of the measured multispectral correlation function, and the definition of the time-averaged intensity, as available due to a finite detector response time.
By combining the two equations, it is immediately recognized that a theoretical evaluation of the correlation function requires computing the fourth-order correlation function of the electric field,
\begin{equation}
    \left\langle
        E^*_a E^*_b E_b E_a
    \right\rangle = 
    \left\langle
        E^*_a(\br_a,t_a)\, E^*_b(\br_b,t_b)\, E_b(\br_b,t_b)\, E_a(\br_a,t_a)
    \right\rangle.
\end{equation}
Its expression can be reduced to a much simpler form by assuming a Gaussian behavior \cite{mandel1995optical} of the stochastic field, so that
\begin{equation}
    \left\langle
        E^*_a E^*_b E_b E_a
    \right\rangle=\left\lvert
        \left\langle
        E^*_b E_a
    \right\rangle\right\rvert^2
       +
        \left\langle
        \left\lvert E_a\right\rvert^2
    \right\rangle
     \left\langle
        \left\lvert E_b\right\rvert^2
    \right\rangle.
\end{equation}
By plugging this simplified form of the fourth-order correlation function into Eq.~(1), one notices that the correlation between the intensity fluctuations reduces to the simple expression
\begin{equation}
    \Gamma(\br_a,\br_b, \Delta t)=
    \frac 1{\tau_\text{exp}^2}
    \int_t^{t+\tau_\text{exp}}\int_{t+\Delta t}^{t+\Delta t+\tau_\text{exp}}\left\lvert
        \left\langle
        E^*_b(\br_b,z_b,t_b) E_a(\br_a,z_a,t_a)
    \right\rangle\right\rvert^2 dt_a\,dt_b.
    \label{eq:coherent}
\end{equation}
The relationship between \eqref{eq:coherent} and the spatial and spectral properties of the sample is found by considering that the electric field at the detection planes $E_{a,b}$, placed at an optical distance $z_{a,b}$ from the source, is the result of the \textit{deterministic} propagation up to the detector planes of a stochastic emission which happening at the source plane. Hence, we can write
\begin{equation}
    E_i(\br_i,z_i,t)=\int\left(
        \int \tilde E_0(\br_0,\omega)\tilde g_i(\br_0,\br_i,\omega)\,d\br_0
    \right) \exp{\left[i\omega\left(\frac {z_i} c -t\right)\right]}d\omega,
    \label{eq:source_to_det}
\end{equation}
for $i=a,b$.
The expression is obtained by decomposing the electric field into its frequency components and by noticing that each frequency can be propagated, through Fourier optics \cite{goodman2005introduction}, from the source plane to the detection plane, provided the frequency-dependent Green's function $\tilde g_i$ are known.
In this picture, all the statistical properties of the correlation function are enclosed within the 2D complex field amplitude of the electric field at the source plane, here conveniently expressed in terms of its frequency content $\tilde E_0(\br_0,\omega)$. 
By comparison with \eqref{eq:coherent}, the properties of the spectral imaging correlation function are thus entirely defined by the first-order degree of spatio-temporal coherence of the field emitted by the source:
\begin{equation}
     \left\langle
        E^*_b(\br_b,z_b,t_b) E_a(\br_a,z_a,t_a)
    \right\rangle= 
    \int
        g_b^*\left(\br_0^\prime,\br_b,\frac {z_b}c-t_b^\prime-t_b\right)
        g_a\left(\br_0,\br_a,\frac {z_a}c-t_a^\prime-t_a\right) 
        g^{(1)}(\br_0,t^\prime_a,\br_0^\prime,t^\prime_b)
        \,d\br_0\,d\br_0^\prime\,dt_a^\prime\,dt_b^\prime,
\end{equation}
where $g^{(1)}=\left\langle E_0^*(\br_0,t^\prime_a)E_0(\br_0^\prime,t^\prime_b)\right\rangle$ is the spatio-temporal degree of coherence at the source \cite{mandel1995optical}, and $g_i$ are the \textit{space-time} Green's functions propagating the electric field from the source plane through space and time to the detectors. These quantities are related to the monochromatic fields and and frequency-dependent Green's functions of~\eqref{eq:source_to_det} through an inverse Fourier transform in the frequency domain
\begin{align}
    E_0(\br_0,t)&=\int \tilde E_0(\br_0,\omega)\exp{\left(-i\omega t\right)}\,d\omega\\
    g_i(\br_0,\br_i,t)&=\int \tilde g_i(\br_0,\br_i,\omega_i)\exp{\left(-i\omega_i t\right)}\,d\omega_i.
\end{align}

The expression of \eqref{eq:coherent} can be significantly simplified through a few reasonable assumptions on the optical design of the apparatus and on the statistical behavior of the source.
Firstly, we shall assume that the coherence length of the physical phenomena leading to the emission is point-like, so that the field at the source can be assumed to show negligible spatial coherence. 
This approximation limits the validity of our model to \textit{local} statistical phenomena. However, one must consider that, in the context of spectral \textit{imaging}, which involves investigating the sample through an imaging device, this assumption is safely satisfied when the scale of the dynamics of emission phenomena is smaller than the resolution of the imaging system, and not only when phenomena are very short range (\textit{e.g.}, at the atomic scale).
Since a typical optical resolution, even in microscopy-oriented schemes, is hardly much smaller that a micrometer, such assumption describes many cases of interest.
In the hypothesis of local emission, the spatio-temporal degree of coherence is entirely defined by the \textit{temporal} degree of coherence,
\begin{align}
    g^{(1)}(\br_0,t^\prime_a,\br_0^\prime,t^\prime_b)
    &\sim \delta^{(2)}(\br_0-\br_0^\prime)\,
    g^{(1)}(\br_0,t^\prime_a-t^\prime_b) \nonumber \\
    &=\delta^{(2)}(\br_0-\br_0^\prime)
    \int \tilde I_0(\br_o,\omega)\exp{\left[-i\omega(t^\prime_a-t^\prime_b)\right]}d\omega.
    \label{eq:g1}
\end{align}
where $\tilde I_0(\br_0,\omega)$, is the power spectral density radiated by each (statistically independent) source point at coordinate $\br_0$; the spectral cube $\tilde I_0$ is thus the quantity ideally measured in a (hyper-)spectral imaging experiment with perfect spatial and spectral resolution.

Let us assume, for definiteness, that the optical design of arm $A$ is such that the final image shows negligible chromatic aberration in the spectral region of interest, namely
\begin{equation}
    \frac{\partial \tilde g_a}{\partial\omega}\simeq0;
\end{equation}
also, let us assume that the spectral arm does not carry any strong dependence on the emission coordinate as is the case in conventional spectroscopy
\begin{equation}
    \frac{\partial \tilde g_b}{\partial \br_0}\simeq0.
    \label{eq:spect_independ}
\end{equation}
Equation~\eqref{eq:coherent} thus becomes
\begin{equation}
    \Gamma(\br_a,\br_b, \Delta t)=
    \frac 1{\tau_\text{exp}^2}
    \int_t^{t+\tau_\text{exp}}\int_{t+\Delta t}^{t+\Delta t+\tau_\text{exp}}
    \left\lvert
    \gamma(\br_a,\br_b, t_b-t_a)
    \right\rvert^2
    \,dt_a\,dt_b,
    \label{eq:realistic}
\end{equation}
with
\begin{equation}
    \gamma(\br_a,\br_b, t_b-t_a)=
    \int
        \tilde I^\prime_0(\br_a,\omega)\, \tilde g_b^*(\br_b,\omega)
        \exp{\left[-i\omega\left(\frac{z_b-z_a}c-(t_b-t_a)\right)\right]}\,d\omega
        \label{eq:gamma}
\end{equation}
where $\tilde I^\prime_0(\br_b,\omega)=\int \tilde I_0(\br_0,\omega)g_a(\br_0,\br_a,\omega)d\br_0$ is the power spectral density, as available with the optical resolution defined by the PSF of the imaging system.
To easily interpret \eqref{eq:realistic}, one can consider the case of isochronous detection ($\Delta t = (z_a-z_b)/c$) with ideal detectors ($\tau_\text{exp}\rightarrow 0$), ideal imaging system $g_a(\br_0,\br_a)\sim\delta^{(2)}(\br_0-\br_a)$, and ideal spectrometer $g_b(\omega,\tilde\omega(\br_b))\sim\delta(\omega-\tilde\omega(\br_b))$, to obtain
\begin{equation}
    \Gamma\left(\br_a,\br_b,\frac{z_b-z_a}c\right)=\left\lvert\gamma_0\left(\br_a,\br_b,\frac{z_b-z_a}c\right)\right\rvert^2\sim 
    \left\lvert 
    \tilde I_0(\br_a,\tilde\omega(\rho_b))
    \right\rvert^2.
    \label{eq:cube}
\end{equation}
In ideal conditions, our technique is, thus, sensitive to the \textit{squared} power spectral density of the sample of interest.
\subsection*{Decomposition of the overall spectral power}
Studying the complete form of \eqref{eq:realistic} unveils other important features of the correlation-based approach.
As we have stated in the main text, multispectral imaging is partucilarly useful when spectrally and spatially identifying \textit{narrow-band} emissions, originating from interesting statistical phenomena, \textit{e.g.} from contaminants. 
Such narrow-band emissions typically share the same spectral region of less relevant broadband phenomena, such as thermal radiation, whose features are known and a source of disturbance within the spectral range they occupy.
As we have reported in the main text, the detectors exposure time can be set to make the correlation function independent of broadband emissions, so as to operate an intrinsic \textit{statistical filtering} that, to the best of our knowledge, has no counterpart in any other spectroscopic technique.
Let us consider that the overall spectrum is contributed by a single broadband phenomenon and $N$ narrow-band emissions, all statistically independent from one another.
The electric field in proximity of the sample can thus be decomposed as a linear superposition of the emission from $N$ narrow-band statistical phenomena $E_{1,\dots,N}$, with bandwidths $\Delta\nu_{1\dots,N}$, and one broadband emission $E_0$, with bandwidth $\Delta\nu_0\gg\Delta\nu_{1,\dots,N}$.
Because of the statistical independence, the first-order degree of coherence $g^{(1)}$ (see \eqref{eq:g1}) reduces to
\begin{equation}
    g^{(1)}(\br_0,\Delta t)=\sum_{n=0}^N\int \tilde I_n(\br_0,\omega)\exp{\left(-i\omega\Delta t\right)}d\omega,
\end{equation}
where $I_n$ is the spectral power corresponding to each statistical contribution.
This results in a \textit{coherent} superposition of the contributions from each independent emission in the correlation function
\begin{equation}
    \Gamma(\br_a,\br_b,\Delta t)=
    \frac 1{\tau_\text{exp}^2}
    \int_t^{t+\tau_\text{exp}}\int_{t+\Delta t}^{t+\Delta t+\tau_\text{exp}}
    \left\{
        \sum_{n=0}^N\left[ \left\lvert\gamma_n\right\rvert^2
        +2 \sum_{m\neq n}\,\text{Re}\left[
        \gamma_n^*\gamma_m
        \right]
        \right]
    \right\}\,dt_a\,dt_b,
    \label{eq:multiEmission}
\end{equation}
where $\gamma_n(\br_a,\br_b, t_b-t_a)$ is the contribution to the correlation function from a single linewidth, defined in analogy to \eqref{eq:gamma}.

\section{Assumptions and parameters for the simulations}
In order to obtain analytical expressions from \eqref{eq:multiEmission} and perform simulations, we have approximated the exposure window of the detectors as a Gaussian function of width $\tau_\text{exp}$.
The PSF of the spectral arm has also been approximated by a Gaussian function
\begin{equation}
    \tilde g_b(\br_b,\omega)\simeq \exp{
    \left[
        -\frac{(\omega-\tilde\omega(\rho_b))^2}{2\Delta\nu_\text{spec}^2}
    \right]},
    \label{eq:Gauss_PSF}
\end{equation}
where $\tilde\omega(\rho_b)$ is correspondence between optical frequencies and detector coordinates defined by the spectral dispersion of the spectrometer and $\Delta\nu_\text{spec}$ is the spectrometer frequency resolution.
Section \ref{sec:Greens} reports calculations for a fiber-coupled spectrometer arm, showing that this approximation is realistic for typical use cases.\\
For simplicity, also the statistical behavior of the emissions has been approximated by Gaussian spectral density functions centered around the $N+1$ peak frequency $\omega_{0,1,\dots,N}$:
\begin{equation}
    \tilde I_n(\br_0,\omega)=I_n(\br_0)\exp{\left[
        -\frac{(\omega-\omega_n)^2}{2\Delta\nu_n^2}
    \right]},
    \label{eq:spectral_density}
\end{equation}
where the function $I_n(\br_0)$ represents the average spatial intensity distribution of the $n$-th linewidth centered around $\omega_n$.
Assuming that the $N$ narrow-band emissions have negligible spectral overlap (Re$\left[\gamma_n^*\gamma_{m\neq n}\right]\simeq0$ for all $n,m=1,\dots,N$), the correlation function reads
\begin{multline}
     \Gamma(\br_a,\br_b,\Delta t)\simeq\sum_{n=0}^N
     \frac{\left\lvert I_n^\prime(\br_a)\right\rvert^2\,\Delta\nu_n^2\,
     \exp{\left[-\frac{\left(\frac{z_b-z_a}c-\Delta t\right)^2}{     \frac1{\Delta\nu_\text{spec}^{2}}+\frac1{\Delta\nu_n^{2}}+
     4\tau_\text{exp}}\right]}
     }{
     \sqrt{1+\frac{\Delta\nu_n^2}{\Delta\nu_\text{spec}^{2}}}  \sqrt{1+\frac{\Delta\nu_n^2}{\Delta\nu_\text{spec}^{2}}+4\tau_\text{exp}^2\Delta\nu_n^2}}
     \exp{\left[
        -\frac{(\tilde\omega(\rho_b)-\omega_n)^2}{\Delta\nu_n^{2}+\Delta\nu_\text{spec}^{2}}
     \right]}
     \\
     +2\sum_{n=1}^N
     \frac{\text{Re}\left[(I_n^\prime(\br_a))^*\,I_0^\prime(\br_a)\right]\,\Delta\nu_n\Delta\nu_0\,
    \exp{\left[-\frac{\left(\frac{z_b-z_a}c-\Delta t\right)^2}{2\left\{
    2\tau_\text{exp}^2+
    \left[
         \left(
     \frac1{\Delta\nu_\text{spec}^{2}}+\frac1{\Delta\nu_n^{2}}
     \right)^{-1}+
     \left(
     \frac1{\Delta\nu_\text{spec}^{2}}+\frac1{\Delta\nu_0^{2}}
     \right)^{-1}
    \right]^{-1}
    \right\}}
    \right]}
     }{     
     \sqrt{1+\frac{\Delta\nu_0^{2}}{\Delta\nu_\text{spec}^{2}}}
     \sqrt{1+\frac{\Delta\nu_n^{2}}{\Delta\nu_\text{spec}^{2}}}
     \sqrt{1+2\tau_\text{exp}^2\left[
     \left(
     \frac1{\Delta\nu_\text{spec}^{2}}+\frac1{\Delta\nu_n^{2}}
     \right)^{-1}+
     \left(
     \frac1{\Delta\nu_\text{spec}^{2}}+\frac1{\Delta\nu_0^{2}}
     \right)^{-1}
     \right]}
     }f_n(\rho_b),
     \label{eq:approx_corrFun}
\end{multline}
where $0\le f_n(\rho_b)\le 1$ is a Gaussian function.
The analysis leading to Fig. 2 of the main text has been obtained by limiting the evaluation of the correlation function to a  spatial point emitting photons contributing to the broadband spectrum $\Delta\nu_0$ and two a narrowband emissions $\Delta\nu_{1,2}$. For clarity, let us write the spectral densities of \eqref{eq:spectral_density} as
\begin{equation}
    \tilde I_{i}(\omega)=\frac{\hbar\,\omega\, n_{i}}{\Delta\nu_{i}}
    \exp{\left[
        -\frac{(\omega-\omega_{i})^2}{2\Delta\nu_{i}^2}
    \right]},
\end{equation}
with $i=0,1,2$; $n_i$ is the photon flux emitted by the three statistical processes, expressed in photons per time unit.
A physical regime in which the measured spectral width of the narrow-band emissions is defined by the spectrometer resolution ($\Delta\nu_{1,2}\ll\Delta\nu_\text{spec}$), whereas the broadband emission is characterized by a width that is much larger than the spectral resolution.
This scenario corresponds to the usual situation in which one is not interested in retrieving the spectral shape of the narrowband phenomena, which would require $\Delta\nu_{1,2}>\Delta\nu_\text{spec}$, but rather to ascertain and resolve the presence and relative intensity of narrowband emissions.
In these conditions, the cross-term Re$\left[\gamma_0^*\gamma_{i}\right]\simeq0$ becomes negligible so that, for isochronous detection $\Delta t=(z_b-z_a)/c$, the correlation function is the sum of a roughly constant offset, defined by the flat profile of the broadband emission, and the spectral contribution of the narrowband emission, whose width is defined by the resolution
\begin{equation}
	\Gamma(\rho_b)\simeq
	[\hbar \tilde\omega(\rho_b)]^2
	\left\{
	a_0	
	+	
	 a_i\,\exp{\left[\frac{(\tilde\omega(\rho_b)-\omega_i)^2}{\Delta\nu_\text{spec}^2}
     \right]}		
	\right\}
\end{equation}
with
\begin{align}
	a_0&=\frac{n_0^2}{
     \sqrt{1+\frac{\Delta\nu_0^2}{\Delta\nu_\text{spec}^{2}}}  \sqrt{1+\frac{\Delta\nu_0^2}{\Delta\nu_\text{spec}^{2}}+4\tau_\text{exp}^2\Delta\nu_0^2}}\\
     a_i&=\frac{n_i^2
	 }{
     \sqrt{1+\frac{\Delta\nu_i^2}{\Delta\nu_\text{spec}^{2}}}  \sqrt{1+\frac{\Delta\nu_i^2}{\Delta\nu_\text{spec}^{2}}+4\tau_\text{exp}^2\Delta\nu_i^2}},
\end{align}
i=1,2.
The visibility of the interesting spectral features above background is thus defined by the contrast $C_\tau=a_i/a_0$. Such contrast is an increasing function of the exposure time, so that at $\tau_\text{exp}=0$ the lowest possible contrast is obtained for
\begin{equation}
C_0 = 
	\left[\frac{a_i}{a_0}\right]_{\tau_\text{exp}=0}\simeq \left(\frac{n_i}{n_0}\right)^2
	\left(\frac{\Delta\nu_0}{\Delta\nu_\text{spec}}\right)^2.
\end{equation}
The ideal-detector contrast, however, can be improved by orders of magnitude by increasing the exposure time, so that for a large exposure time one gets
\begin{equation}
C_\infty=
	\left[\frac{a_i}{a_0}\right]_{\tau_\text{exp}\rightarrow\infty}\simeq \left(\frac{n_i}{n_0}\right)^2
	\frac{\Delta\nu_0}{\Delta\nu_\text{spec}}
	\frac{\Delta\nu_0}{\Delta\nu_i},
\end{equation}
so that the contrast can be improved by up to a factor $\Delta\nu_\text{spec}/\Delta\nu_i$, as discussed in the main text.

Fig. 3 of the main text has been obtained by considering a sample composed of two narrow-band emitters from separated regions in space.
The areas of the sample from which the emission occurs are not perfectly in focus on the imaging device, which thus collects a \textit{defocused} spectral image, so that the correlation function of \eqref{eq:realistic} is a coherent superposition of two contributions $\gamma=\gamma_1+\gamma_2$, with
\begin{equation}
    \gamma_i(x_a,\,x_b,\,t_b-t_a) = \int \tilde{I}_i(x_0,\omega)\tilde g^*_{a,i}(x_0,x_a)\,
    \tilde g_b(x_b,\omega)\exp{\left[-i\omega\left(\frac{z_{b}-z_{a}}c-(t_b-t_a)\right)\right]}dx_0\,d\omega,
\end{equation}
where $i=1,2$, $\tilde g_{a,i}$ are the two imaging-arm Green's functions, corresponding to the two different axial placements of the emitters, $\tilde g_b$ is supposed independent of the axial position, and the phase delay due to defocusing is supposed negligible compared to the total phase delay through the optical path ($\exp[i\omega(z_{a,i}-z_{b,i})/c]\simeq\exp[i\omega(z_{a}-z_{b})/c]$). Furthermore, the $y$-components of 2D vectors are neglected for brevity, assuming the only relevant spatial and spectral features are distributed along the $x$-dimension.
Since there is no broadband emission to suppress, we can assume an ideal detector time response, so that the correlation function reduces, for isochronous detection, to
\begin{equation}
    \Gamma(x_a,\tilde\omega(x_b))=
    \left\lvert
        \int\left[ \sum_{n=1,2}
        \int\tilde{I}_n(x_0,\omega)
        g^*_{a,n}(x_0,x_a)dx_0\right]\,
    \tilde g_b(x_b,\omega)\,d\omega
    \right\rvert^2.
    \label{eq:corr_comparison}
\end{equation}
From the equation, we see that the resulting image on the spatial arm is a coherent superposition of the images at the various emission frequencies.

\section{Green's function of the spectrometer}\label{sec:Greens}

For the sake of showing that the assumptions we made in the main text for the properties of the proposed spectral imaging device, we show the calculation for a typical Green's function of a spectrometer $g_b(\br_0,\br_b,\omega)$. We assume that, after entering the first optical element of the device, the electric field emitted by the object is divided by a beam-splitter into the two optical arms.
While arm A is a conventional imaging device, which only affects the device with the resolution it defines, the properties of the spectral arm are more relevant to this work and will be analyzed here.
\begin{figure}
    \centering
    \includegraphics[width=0.8\textwidth]{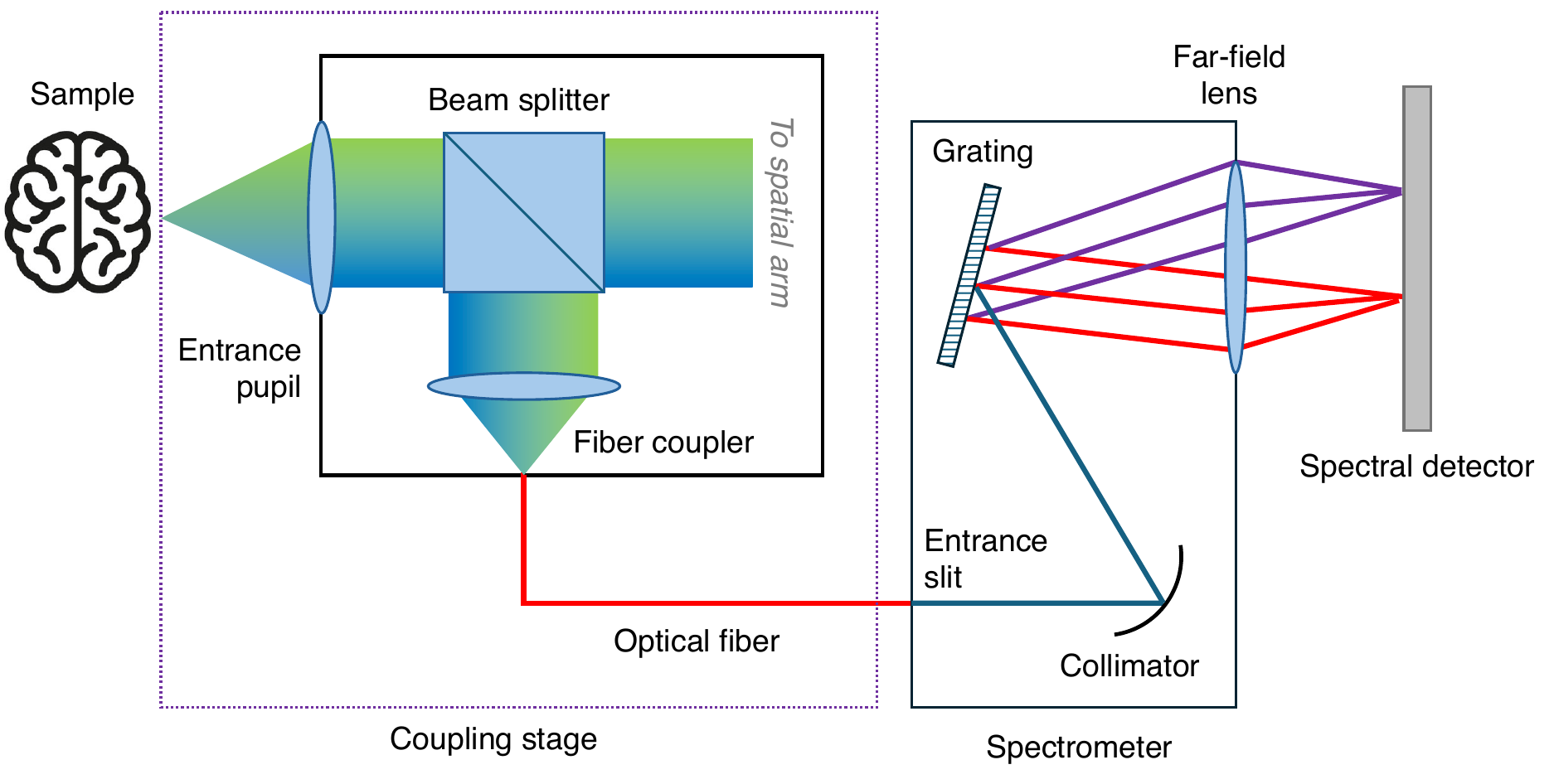}
    \caption{Schematics of the spectral arm}
    \label{fig:scheme}
\end{figure}
The spectral image, which combines spectrum and intensity distribution, is obtained by evaluating the cross-correlation function between light intensity \textit{fluctuations} at the two detectors
\begin{equation}
    \Gamma(\br_a,\br_b, \Delta t)=\left\langle
        \Delta\bar{I}_a(\br_a, t)\,\Delta\bar{I}_b(\br_b, t+\Delta t)
    \right\rangle
    \label{eq:corrFun_definition}
\end{equation}
After the beam splitter, let us assume the electric field is coupled into an optical fiber and transferred to the entrance slit of a spectrometer, with negligible size.
The operation of coupling the field into a non-image-preserving fiber is what makes so that the the dependence of the Green's function on the emission coordinate $\br_0$ is lost, as assumed in \eqref{eq:spect_independ}. We shall assume that the chromatic dispersion happens along the $x$-axis of the system, so that the other dimension shall be neglected ($\br\rightarrow x$)
The finite angular acceptance of the imaging device and the coupling efficiency reflect into a complex coefficient $c(\omega)$, whose amplitude determines frequency-dependent coupling efficiency and whose phase can change throughout the spectrum.
After entering the spectrometer, light is collimated and sent to a diffraction grating for spectral separation. The dispersed electric field is then imaged by a lens onto the spectral detector $D_b$, placed in its second focal plane.
For ease of calculation, we shall consider only the $m$-th diffraction order from the grating. This situation corresponds, for instance, to the case of a blazed diffraction grating, where most of the light is diffracted into a single order, selected to optimize the free spectral range (FSR) and amount of chromatic dispersion. With this assumptions, the blazed grating can be approximated as a transmissive mask
\begin{equation}
	\exp{(4\pi i\,m\,\nu\,x_g)},
\end{equation}
where $\nu$ is the spatial frequency of the diffraction grating, and $x_g$ is the coordinate on the grating plane.
We shall assume that the main limitation to the wavefronts, defining the spectral resolution, in combination with the spectral resolving power of the grating, happens at the level of the collimation lens, having an acceptance pupil of diameter $l$.
For simplicity, we also assume the focal lengths of the collimation and final lens to be the same and equal to $f$.\\
The Green's function reads, up to irrelevant constants
\begin{equation}
	g_b(x_b,\omega)=\omega\,c(\omega)\,
	\exp{\left[
	\left(
			-i\frac{x_b}{\omega\,l}
		\right)
		\frac{(\omega-\tilde\omega(x_b))^2}{2 \Delta\nu_\text{spec}^2}
	\right]}
	\text{sinc}\left[
		\frac{\omega-\tilde\omega(x_b)}{\Delta\nu_\text{spec}}	
	\right],
\end{equation}
with $\tilde\omega(x_b)=4\pi m\frac{c\,f}{x_b}\nu$ defined by the grating dispersion, and $\Delta\nu_\text{spec}=\frac{c\,f}{x_b\,l}$ mainly defined by the system numerical aperture ($l/f$).
Assuming the FSR is small compared to the typical variations of the coupling efficiency $c(\omega)$ and to the frequency $\omega$, which is always true for optical frequencies, then the multiplying function $\omega\,c(\omega)$ can be approximated to a good extent as a constant.
Furthermore, the complex phase in the Green's function is proportional to the coefficient $\frac{x_b/l}{\omega}$. Considering that, for paraxial propagation, the numerator is small, such coefficient is zero at optical frequencies.
Up to irrelevant complex multiplying constants, the spectrometer Green's function is a real function peaked around $\tilde\omega(x_b)$ and $\Delta\nu_\text{spec}$ wide.
If Gaussian apodization is assumed, the function reduces to the form of \eqref{eq:Gauss_PSF}.

\end{document}